# Coherent magnetization precession in ferromagnetic (Ga,Mn)As induced by picosecond acoustic pulses

A.V. Scherbakov[1], A.S. Salasyuk[1,2], A.V. Akimov[1,3], X. Liu[4], M. Bombeck[2], C. Brüggemann[2], D.R. Yakovlev[1,2], V.F. Sapega[1], J.K. Furdyna[4], and M. Bayer[2]

[1]*Ioffe Physical-Technical Institute of the Russian Academy of Sciences, 194021 St. Petersburg, Russia*

[2]*Experimentelle Physik 2, Technische Universität Dortmund, D-44227 Dortmund, Germany*

[3]*School of Physics and Astronomy, University of Nottingham, Nottingham NG7 2RD, UK*

[4]*Department of Physics, University of Notre Dame, Notre Dame, Indiana 46556, USA*

**Abstract**

We show that the magnetization of a thin ferromagnetic (Ga,Mn)As layer can be modulated by picosecond acoustic pulses. In this approach a picosecond strain pulse injected into the structure induces a tilt of the magnetization vector **M**, followed by the precession of **M** around its equilibrium orientation. This effect can be understood in terms of changes in magneto-crystalline anisotropy induced by the pulse. A model where only one anisotropy constant is affected by the strain pulse provides a good description of the observed time-dependent response.

The success of semiconductors in today's modern technology arises from our ability to tailor their electrical and optical properties on a detailed level. High crystal quality allows one to implement tools for fast manipulation of properties, which is of interest both for fundamental studies and for applications. In this respect, manipulation by high-frequency ($10^9$-$10^{12}$ Hz) acoustic waves has become increasingly attractive for extending traditional acousto-electronics and acousto-optics to gigahertz (GHz) and terahertz (THz) frequency ranges [1]. During the last decade intense efforts have been undertaken to extend the application spectrum by making semiconductors also magnetic. Therefore one might also seek for ultrafast control of semiconductor magnetism by high-frequency sound. Access to such control could be obtained through the strong sensitivity of the magneto-crystalline anisotropy (MCA) in ferromagnetic semiconductors, such as (Ga,Mn)As to strain [2-4].

Several methods to control MCA in (Ga,Mn)As films by applying an electric field and varying doping have already been developed [5-9]. In ultrafast experiments, modulation of MCA was realized by optically-induced increases of carrier density and lattice temperature [10-12]; and coherent control of the magnetization precession [13] as well as femtosecond switching of magnetization [14,15] was demonstrated. The application of ultrafast optical methods, however, is limited by effects such as simultaneous generation of large numbers of nonequilibrium carriers and phonons. The control of MCA by acoustic waves would allow ultrafast manipulations of magnetization in a ferromagnetic semiconductor without these side-effects.

The goal of the present work is to inject an ultrashort high-amplitude acoustic wavepacket into a ferromagnetic (Ga,Mn)As layer, and to monitor the relevant magnetization changes. We use the methods of ultrafast acoustics, which enable the generation of picosecond strain pulses in solids [16]. Strain pulses with the amplitude up to $10^{-3}$ generated by femtosecond optical pulses in thin metal films and injected into crystalline substrates (e.g. GaAs) have been shown to travel over millimeter distances at low temperatures, and up to 100 μm at room temperature [17]. Such a strain pulse has a direct, short and intense impact on the MCA of the layer, and thus can act as an instrument for magnetization control.

In this letter we show that strain pulses injected into a ferromagnetic (Ga,Mn)As layer induce a tilt of magnetization from its stationary equilibrium orientation, followed by coherent precession of the magnetization at a frequency of ~10 GHz. This process can be well described by a simple model where the strain pulse induces a change in the MCA of (Ga,Mn)As.

The sample used in our studies consists of a single $Ga_{0.95}Mn_{0.05}As$ layer with a thickness $d$=200 nm grown by low-temperature molecular beam epitaxy on a semi-insulating (001) GaAs substrate. SQUID magnetometery shows that the Curie temperature of the sample is 60 K and the saturation magnetization $M$ is 20 emu/cm$^3$. The magnetic layer is under an in-plane compressive strain that leads to an in-plane orientation of easy magnetization axis [3]. The values of the internal strain components obtained from X-ray diffraction are $-2\times10^{-3}$ for the in-plane compressive strain, and $1.9\times10^{-3}$ for the out-of-plane tensile strain. In the absence of an external magnetic field the spontaneous magnetization **M** of magnetic domains lies in the layer plane at low temperatures. The magnetic field **B** applied along the $z$-axis (i.e., perpendicular to the layer) turns **M** out of the layer plane, as shown in Fig. 1(a). We will describe the direction of **M** in a single magnetic domain by the angles $\theta$ and $\psi$ as defined in Fig. 1(a).

Our experiments were carried out at $T$=1.6 K in a cryostat equipped with a superconducting magnet. The projection of **M** on the $z$-axis, $M_z=M\cos\theta$, is measured by monitoring the Kerr rotation (KR), i.e., the angle $\varphi$ between the linear polarizations of the light incident on and reflected from the (Ga,Mn)As layer, given by:

$$\varphi = \arctan\left(\text{Re}\left[i\frac{r^+ - r^-}{r^+ + r^-}\right]\right), \qquad (1)$$

where $r^+$ and $r^-$ are complex reflection coefficients for left- and right-handed circularly polarized light. In general a difference between $r^+$ and $r^-$ may arise from several causes. In (Ga,Mn)As layers KR has a contribution from the magneto-optical Kerr effect [18] and in our experimental geometry the KR angle $\varphi$ is proportional to the **M** component along z-axis, $M_z$, i.e. $\varphi \propto M_z$.

Figure 1(b) shows the magnetic field dependence of the angle $\varphi$ and the corresponding value of $M_z/M=\cos\theta$ at stationary equilibrium conditions. The measured field dependence of these quantities is typical for (Ga,Mn)As layers with the easy magnetization axis in the (001) plane [17,18]. Note that the value of $\varphi$ – and thus of $M_z$ – increases with B and finally saturates at $B>2$ kOe, i.e., when **M**||**B** and $M_z=M$. The dip/kick observed around $B=1$ kOe is known to be caused by optical interference effects [19]. The field dependence of the magnetization can be analyzed by minimizing the free energy with respect to $\theta$, as described in Ref. [20], allowing us to obtain the anisotropy parameters $H_{2\perp}$, $H_{4\perp}$, $H_{2\|}$ and $H_{4\|}$ of the sample, i.e., the perpendicular uniaxial, perpendicular cubic, in-plane uniaxial, and in-plane cubic anisotropy fields, respectively. The following values of anisotropy fields provide the best fit of the experimental curve: $(4\pi M + H_{2\perp}) = 1.82$ kOe, $H_{4\perp} = 0.66$ kOe, $H_{2\|} = 0$ and $H_{4\|} = 1.97$ kOe. These values are typical for ferromagnetic (Ga,Mn)As layers with Mn content of *ca.* 5% and with in-plane easy magnetization axes along [100] or [010] directions [20].

In the ultrafast acoustic experiments [Fig. 1(c)] we generate strain pulses, which modify the MCA, thus, causing the magnetization to turn from its equilibrium position, and monitor in real time the resulting changes in $M_z$. Picosecond strain pulses are generated in a 100-nm thick Al film deposited on the back side of the GaAs substrate. The film is excited by optical pump pulses from an amplified femtosecond laser. The calculated [21] spatial shape of the strain pulse $\varepsilon_{zz}(z,t)$ injected into the substrate is shown in Fig. 1(d). The pulse propagates in GaAs with a longitudinal sound velocity $v=4.8$ km/s, and in time $t_0=l_0/v \approx 22$ ns (where $l_0=105$ μm is the substrate thickness) reaches the magnetic layer. At the open surface of the film the strain pulse is reflected with a phase inversion, and travels back towards the GaAs substrate. Thus the thickness $d$ of the (Ga,Mn)As layer is modulated by the strain pulse. The temporal profile of the relative layer thickness modulation can be written as:

$$\varepsilon(t) = \frac{\Delta d(t)}{d} = \frac{1}{d}\int_0^d \varepsilon_{zz}(t,z)dz \;, \qquad (2)$$

where $\Delta d(t)$ is the time evolution of the film thickness change, and $z=0$ corresponds to the interface between the GaAs substrate and (Ga,Mn)As film. The time evolution of $\varepsilon(t)$ for our experimental conditions is shown in Fig. 1(e) by the solid line. It is seen that the strain pulse results in compressive and tensile perturbations of the film, separated by intervals when $\Delta d=0$.

The time evolution of the magnetization is monitored by probing $M_z$ with sub-picosecond time resolution. We measure the strain-induced modulation of the KR angle $\varphi$ as a function of the time delay $t$ between pump and probe pulses. The optical probe pulse, split from the same laser beam, is focused to a spot on the (Ga,Mn)As layer opposite to the pump excitation [Fig.1(c)]. Figure 2(a) shows the time evolution of the KR angle change $\Delta\varphi(t)$ measured at different $B$, where $t=0$ corresponds to the time when the strain pulse reaches the (Ga,Mn)As layer. In the time interval from $t=0$ to $t=125$ ps indicated in Fig. 2(a) by the vertical arrow the strain pulse travels through the (Ga,Mn)As layer to the edge of the sample and back to the GaAs substrate, so that for $t>125$ ps the acoustic wavepacket is no more present in the (Ga,Mn)As layer . The remarkable experimental result is that in the field range $0<B<2$ kOe and at $t>150$ ps i.e., *after* the strain pulse itself has already left the ferromagnetic layer, the KR signal shows pronounced oscillations with a frequency ~10 GHz,. These low-frequency oscillations are not detected at $B=0$. They only appear when $B$ is applied; and they disappear at high magnetic field above 2.5 kOe. These oscillations last for times ~1 ns and their frequency and amplitude depend on $B$. As an example, the KR signal obtained at $B=0.8$ kOe, where the low frequency oscillations have the highest amplitude, is shown in Fig. 2(b).

In addition to these low frequency oscillations, $\Delta\varphi(t)$ also reveals more complex fast oscillating features. Specifically, high frequency oscillations of about 44 GHz are observed at $B>1$ kOe in a wide time interval. Oscillations of this type have been seen earlier in GaAs, (Al,Ga)As and (In,Mn)As films [22-24], and are ascribed to interference of the probe beams reflected from the sample surface and from the strain wave packet propagating in the sample. Magnetic field makes these oscillations evident in the KR due to the circular dichroism of the paramagnetic GaAs substrate when the strain pulse propagates through it at $t<0$ and $t>125$ ps [see Eq. (1)]. There is also a pronounced fast oscillating contribution to $\Delta\varphi(t)$ in the time interval $0<t<125$ ps, i.e. when the strain pulse is traveling in the (Ga,Mn)As layer. This high frequency contribution is present at all fields, including $B=0$, and therefore cannot be attributed to the modulation of magnetization. Apparently this contribution is due to the anisotropy of the elasto-optical constants in the strained (Ga,Mn)As layer, which gives rise to the KR in accord with Eq. (1). Detailed analysis of this effect can be performed in terms of the theoretical approach presented in Refs. [25, 26], but this is beyond the scope of the present work.

In what follows we will concentrate on the long-lived low-frequency oscillations of $\Delta\varphi(t)$, which we attribute to the modulation of $M_z$ in (Ga,Mn)As. We consider these oscillations to be due to the strain-induced tilt of the magnetization vector **M**, followed by a coherent precession of **M** around its equilibrium direction. The frequency of these oscillations is in the GHz range typical for such precession, as established by ferromagnetic resonance experiments [27]. The amplitude of the oscillations is expected to be negligible at $B=0$, where the net z-component of **M** vanishes. At high fields, $B>2$ kOe, **M** is practically parallel to **B** ($\theta\rightarrow 0$), and the temporal modulation of $M_z$ also becomes negligibly small. Thus the strain-induced tilt of the magnetization and its precession should be most conspicuous in the range of magnetic field where the direction of **M** is determined by the balance between the external magnetic field and the MCA field. All these arguments can be applied to the behavior of $\Delta\varphi(t)$ observed experimentally in the range $0<B<2$ kOe.

We now analyze the magnetization kinetics associated with the strain pulse propagating in the (Ga,Mn)As layer. In the presence of a magnetic field applied normal to the layer, a (Ga,Mn)As film with a thickness of 200 nm has a single value of $M_z$, in that respect behaving like a single domain [28]. We may, therefore, express quantitatively the temporal evolution of magnetic anisotropy in the (Ga,Mn)As layer through the relative change of the layer thickness $\varepsilon(t)$ given by Eq. (2). Since then any modification of the built-in stationary strain produced by $\varepsilon(t)$ will tilt the equilibrium magnetization by an angle $\Delta\theta_\varepsilon(t)$ relative to its unperturbed stationary direction. When $\varepsilon(t)$ is much less than the built-in equilibrium strain, $\Delta\theta_\varepsilon(t)$ is proportional to $\varepsilon(t)$ and we may write:

$$\Delta\theta_\varepsilon(t) = \frac{d\theta}{d\varepsilon_{zz}}\varepsilon(t) = \frac{d\theta}{dH_{2\perp}}\frac{dH_{2\perp}}{d\varepsilon_{zz}}\varepsilon(t) \qquad (3)$$

In Eq. (3) we have assumed that $\varepsilon(t)$ changes only the value of the uniaxial perpendicular term $H_{2\perp}$ of magneto-crystalline anisotropy, and the magnetization is tilted only by the change in the angle $\theta$ [29].

To simplify the analysis, we approximate $\varepsilon(t)$ by square pulses shown in Fig. 1(e) by dashed lines. Then in the time intervals when $\varepsilon(t)$ remains constant **M**(*t*) possesses a circular precession with frequency *F* around the direction defined by $\theta$ and $\Delta\theta_\varepsilon(t)$. Every time when $\varepsilon(t)$ changes to another constant value, **M**(*t*) starts to precess circularly around another, corresponding to $\varepsilon(t)$, direction. With this simplification we seek the values of $d\theta/d\varepsilon_{zz}$ and the precession frequency *F* that give the best agreement between the calculated temporal evolution of $\Delta M_z/M$ and the detected KR signal $\Delta\varphi(t)$ at a given field. An example of the fit $\Delta M_z(t)/M$ for $d\theta/d\varepsilon_{zz}$ =60 radian and *F*=6.86 GHz, which fits well the experimental signal for *B*=0.8 kOe is shown by the solid line in Fig. 2(b). A similar procedure was used for other values of *B*, yielding the corresponding magnetic field dependences of $d\theta/d\varepsilon_{zz}$ and *F*, which are shown by symbols in Fig. 3. The solid curve in Fig. 3(a) shows the field dependence of the frequency *F* calculated using the MAC parameters for our sample (for details see Ref. [20]). The experimental and calculated dependences demonstrate a clear decrease of *F* with increasing *B,* and agree well with each other. The solid curve in Fig. 3(b) is the calculated dependence of $d\theta/d\varepsilon_{zz}$ obtained by minimizing the free energy with respect to $\theta$, assuming a field-independent value of $dH_{2\perp}/d\varepsilon_{zz}$ as a fitting parameter [see Eq. (3)]. Excellent agreement with experimental data is achieved using $dH_{2\perp}/d\varepsilon_{zz}$ = 850 kOe. This value is in good agreement with the data recently reported for several (Ga,Mn)As samples [3]. The dependence of $d\theta/d\varepsilon_{zz}$ on $B$ in Fig. 3(b) shows that it increases slowly for $B<1$ kOe, then quite rapidly for $B>1$ kOe, and at higher fields ($B>2$ kOe, when **M** is almost parallel to **B**) $d\theta/d\varepsilon_{zz}$ eventually falls to 0. Such behavior qualitatively follows the $M_z(B)$ curve in Fig. 1(b), which also shows first a slow, and then a rapid increase in corresponding field ranges. Thus, the response of **M** either to the external magnetic field or to strain-pulse-induced changes in MCA fields shows similar features.

To conclude, we have demonstrated the effect of high-frequency acoustic pulses on the magnetization of a ferromagnetic (Ga,Mn)As layer. The strain-induced modulation of magneto-crystalline anisotropy in (Ga,Mn)As by a subTHz acoustic wavepacket leads to the tilt of magnetization followed by its coherent precession around equilibrium direction. The model based on the assumption that the strain pulse modulates only the single magneto-crystalline anisotropy term $H_{2\perp}$ is in excellent quantitative agreement with the experimental observations. Based on these results, we suggest that control of magnetization by THz acoustic pulses (e.g. shock wave and soliton train [30]) on even faster time scales may be realized experimentally

The authors thank L.E. Golub, N.S. Averkiev and S.A. Tarasenko for valuable discussions. This work was supported by the Deutsche Forschungsgemeinschaft via Koselleck Programme (Grant No. BA1549/14-1), the Russian Foundation for Basic Research, the Russian Academy of Sciences, CRDF (Grant. No. RUP 1-2890-ST-07) and the National Science Foundation (grant DMR06-03762).

References

[1] M. R. Armstrong et al., Nature Physics **5**, 285 (2009).

[2] J. Zemen et al., Phys. Rev. B **80**, 155203 (2009).

[3] M. Glunk et al., Phys. Rev. B **79**, 195206 (2009).

[4] U. Welp et al., Phys. Rev. Lett. **90**, 167206 (2003).

[5] M. Overby et al., Appl. Phys. Lett. **92**, 192501 (2008).

[6] A. W. Rushforth et al., Phys. Rev. B **78**, 085314 (2008).

[7] C. Bihler et al., Phys. Rev. B **78**, 045203 (2008).

[8] Sunjae Chung et al., Solid State Commun. **149**, 1739 (2009).

[9] D. Chiba et al., Nature **455**, 515 (2008).

[10] J. Qi et al., , Appl. Phys. Lett. **91**, 112506 (2007).

[11] Y. Hashimoto, S. Kobayashi, and H. Munekata, Phys. Rev. Lett. **100**, 067202 (2008).

[12] E. Rozkotová et al., Appl. Phys. Lett. **92**, 122507 (2008).

[13] E. Rozkotová et al., Appl. Phys. Lett. **93**, 232505 (2008).

[14] G. V. Astakhov et al., Appl. Phys. Lett. **86**, 152506 (2005).

[15] K. C. Hall et al., Appl. Phys. Lett. **93**, 032504 (2008).

[16] http://en.wikipedia.org/wiki/Picosecond_ultrasonics

[17] H.-Y. Hao and H. J. Maris, Phys. Rev. B **63**, 224301 (2001)

[18] R. Lang et al., Phys. Rev. B **72** 024430 (2005).

[19] G. P. Moore et al., J. Appl. Phys. **94**, 4530 (2003).

[20] X. Liu et al., J. Appl. Phys. **98**, 063904 (2005).

[21] G. Tas and H. J. Maris, Phys. Rev. B **49**, 15046 (1994).

[22] O. B. Wright et al., Phys. Rev. B 64, 081202 (2001)

[23] H. B. Zhao et al., Appl. Phys. Lett. **86**, 152512 (2005).

[24] J. Wang et al., Phys. Rev. B **72**, 153311 (2005).

[25] H.N. Lin et al., J. Appl. Phys. **69**, 3816 (1991).

[26] O. B. Wright, J. Appl. Phys. **71**, 1617 (1992).

[27] X. Liu, Y. Sasaki, and J. K. Furdyna, Phys. Rev. B, **67** 205204 (2003).

[28] T. Shono et al., Appl. Phys. Lett. **77**, 1363 (2000).

[29] Actually, if $\varepsilon(t)$ also has an effect on the angle $\psi$, the in-plane multi-domain structure of (Ga,Mn)As should average out this effect.

[30] H.-Y. Hao and H. J. Maris, Phys. Rev. B **64**, 064302 (2001).

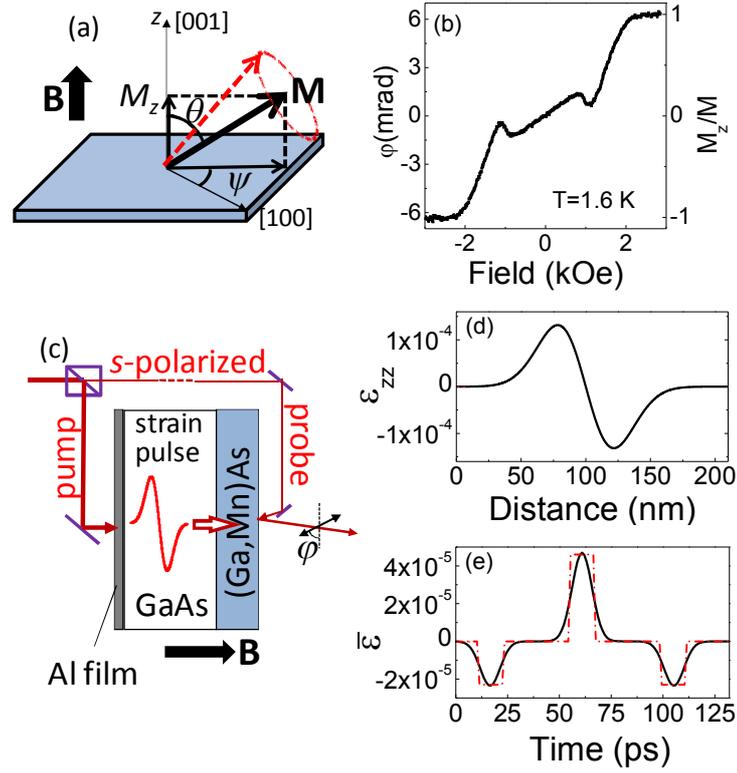

**Figure 1**. (a) Schematic of sample and magnetic field orientations. The dashed circle shows the precession of **M** around its equilibrium direction, which results in a modulation of $M_z$. (b) Magnetic field dependence of the KR angle (left vertical axis) and corresponding value of $M_z/M$ (right vertical axis) measured in the absence of strain pulses. (c) Schematic of pump-probe experiments with strain pulses. The laser characteristics are: wavelength 800 nm, pulse duration 200-fs, repetition rate 100 kHz; pump beam spot diameter is 300-µm and the energy density per pulse is 2mJ/cm$^2$; probe beam spot diameter is 150-µm and energy density per pulse is less than 10 µJ/cm$^2$ .(d) Spatial shape of strain pulse $\varepsilon_{zz}$ injected into the GaAs substrate. (e) Temporal evolution of relative layer thickness of the (Ga,Mn)As layer $\varepsilon(t)$.

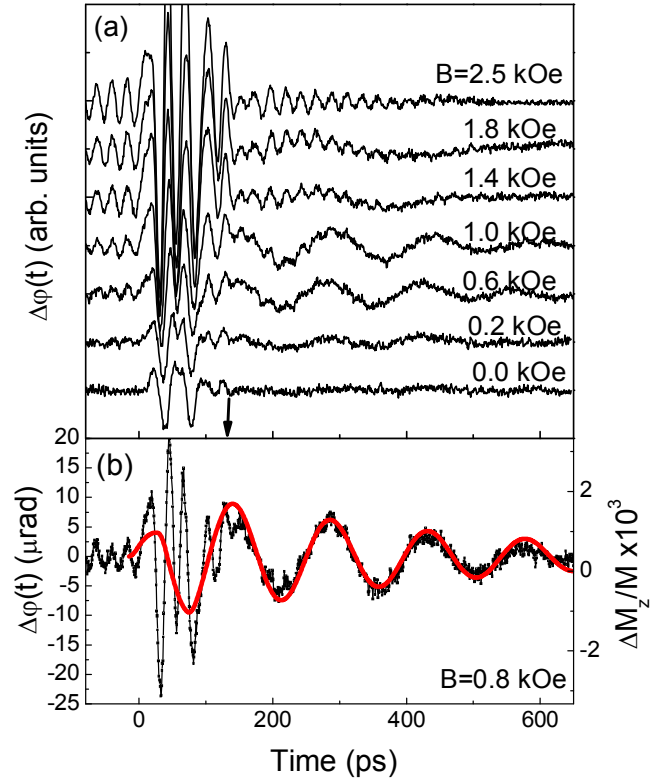

**Figure 2**. Strain pulse-induced temporal evolution $\Delta\varphi(t)$ of the KR angle changes measured (a) at various magnetic fields; and (b) at $B$=0.8 kOe. The value $t$=0 corresponds to the time when the strain pulse enters the (Ga,Mn)As layer, and the vertical arrow shown in panel (a) indicates the time when the strain pulse leaves the (Ga,Mn)As layer. The thick solid curve in (b) is the calculated temporal evolution of $\Delta M_z(t)/M$ obtained for $d\theta/d\varepsilon_{zz}$ =60 rad, $F$=7 GHz and for a precession decay time of 400 ps.

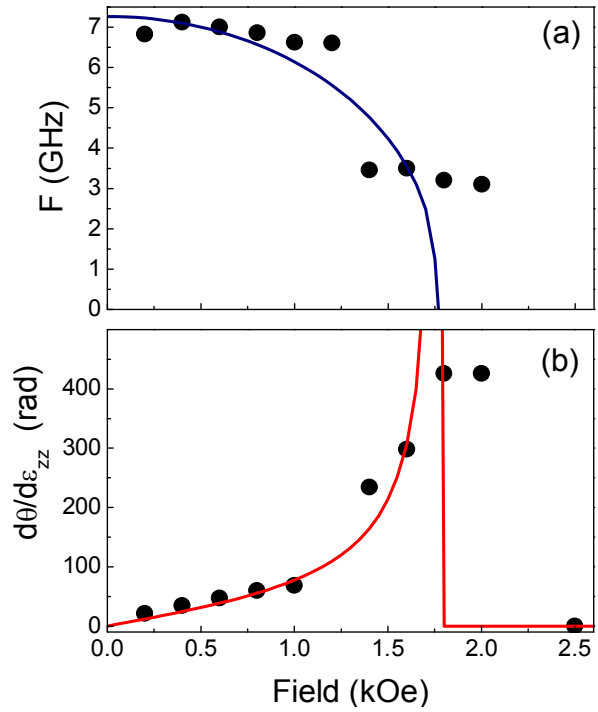

**Figure 3**. Field dependences of (a) precession frequency $F$; and (b) the angle variation parameter $d\theta/d\varepsilon_{zz}$. Symbols show the values obtained by fitting the experimental KR signals. Solid lines show the calculated dependences of $F$ and $d\theta/d\varepsilon_{zz}$ obtained using the anisotropy parameters of the studied structure and assuming $dH_{2\perp}/d\varepsilon_{zz}$ to be a field-independent parameter equal to 850 kOe.